\documentclass[sigconf]{acmart}
\usepackage{graphicx}
\usepackage{latexsym}
\usepackage{multirow}
\usepackage{balance}
\usepackage{paralist}
\usepackage[super]{nth}
\usepackage[compatibility=false,font=bf]{caption}
\usepackage{subcaption}
\usepackage{amsmath}
\usepackage{amssymb}
\usepackage{mathtools}
\usepackage{bm}
\usepackage{amsfonts}
\usepackage[normalem]{ulem}

\usepackage{dblfloatfix}

\usepackage{todonotes}

\usepackage[linesnumbered,algoruled,noend]{algorithm2e}
\SetAlFnt{\small}
\SetAlCapFnt{\small}
\DontPrintSemicolon

\usepackage{multirow}
\usepackage{booktabs}

\usepackage{tikz}
\usepackage{pgfplots}
\usepackage{pgfplotstable} 
\pgfplotsset{compat=1.14}
\usepgfplotslibrary{groupplots}
\usepgfplotslibrary{fillbetween}
\usetikzlibrary{patterns}

\newtheorem{problem}{Problem}

\newcommand{\errorband}[6]{
	\addplot [name path=pluserror,draw=none,no markers,forget plot]
	table [x={#2},y expr=\thisrow{#3}+\thisrow{#4}] {#1};
	
	\addplot [name path=minuserror,draw=none,no markers,forget plot]
	table [x={#2},y expr=\thisrow{#3}-\thisrow{#4}] {#1};
	
	\addplot [forget plot,fill=#5,opacity=#6]
	fill between[on layer={},of=pluserror and minuserror];
}

\renewcommand{\leq}{\leqslant}
\renewcommand{\geq}{\geqslant}

\begin{document}

\copyrightyear{2018} 
\acmYear{2018} 
\setcopyright{acmcopyright}
\acmConference[CIKM '18]{2018 ACM Conference on Information and Knowledge Management}{October 22--26, 2018}{Torino, Italy}
\acmPrice{15.00}
\acmDOI{10.1145/3269206.3269236}
\acmISBN{978-1-4503-6014-2/18/10} 
\acmBooktitle{2018 ACM Conference on Information and Knowledge Management (CIKM'18), October 22--26, 2018, Torino, Italy}

\fancyhead{}

\title{Efficient Taxonomic Similarity Joins with Adaptive Overlap Constraint}


\author{Pengfei Xu and Jiaheng Lu}
\affiliation{Department of Computer Science, University of Helsinki}
\email{first.last@helsinki.fi}

\begin{abstract}
	A similarity join aims to find all similar pairs between two collections of records. Established approaches usually deal with synthetic differences like typos and abbreviations, but neglect the semantic relations between words. Such relations, however, are helpful for obtaining high-quality joining results. In this paper, we leverage the taxonomy knowledge (i.e., a set of IS-A hierarchical relations) to define a similarity measure which finds semantic-similar records from two datasets. Based on this measure, we develop a similarity join algorithm with prefix filtering framework to prune away irrelevant pairs effectively. Our technical contribution here is an algorithm that judiciously selects critical parameters in a prefix filter to maximise its filtering power, supported by an estimation technique and Monte Carlo simulation process. Empirical experiments show that our proposed methods exhibit high efficiency and scalability, outperforming the state-of-art by a large margin.
\end{abstract}



\maketitle





\section{Introduction}
\label{sec:introduction}




Given two sets of records, a similarity join aims to find all records whose similarities are higher than a given threshold. Such operation is widely-seen in tasks such as data cleaning \cite{journals/tods/LuLWLX15,conf/icde/ArasuCK08}, information retrieval \cite{conf/www/BayardoMS07,conf/dasfaa/XuL17}, and data mining \cite{conf/sigmod/LuLWLW13}. To perform joining efficiently, a plethora of established algorithms utilise similarity measures, e.g., \textit{Levenshtein similarity} \cite{conf/sigmod/WangLF12} and \textit{Jaccard coefficient} \cite{conf/sigmod/Sarawagi04}. Such measures capture syntactic-similar records, which is not enough because of the existence of synonyms and related concepts, which often differ from spellings.

\begin{figure}[t]
    \centering
    \includegraphics[width=0.98\linewidth]{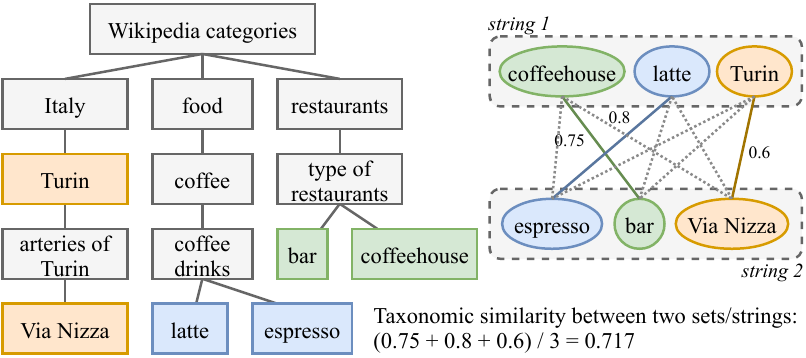}
    \caption{Example of a simplified hierarchical taxonomy and similarity calculation based on Equations \ref{eq:ts_ic} and \ref{eq:gts_ic}.}
    \label{fig:toytaxonomyexample}
    \vspace*{-3mm}
\end{figure}

Taxonomy is an abundant source of lexica, maintaining IS-A relations between terms. It has been proved useful \cite{journals/tkde/ShangLLF16} for enhancing the quality of similarity joins. Figure \ref{fig:toytaxonomyexample} depicts an example. Given a Wikipedia taxonomy and two strings, a join algorithm with Levenshtein distance will fail to capture their similarity due to distinct spellings. In contrast, a \textit{taxonomic similarity measure} maps each string to multiple taxonomy nodes, and calculate the similarity between every two nodes from the depth of their lowest common ancestor (LCA). For example, ``Turin'' and ``Via Nizza'', with three common ancestors (including ``Turin'' itself) and maximal depth five, have 0.6 (=3/5) similarity. Hence, the similarity between two strings can be calculated as 0.717, by averaging the maximal sum of three distinct node-wise similarities.

Joins with taxonomic similarity measure can be useful in many real-life scenarios. For example, location providers are interested in integrating knowledge taxonomy to remove duplicates or link related records from crawled Points of Interests (POIs) \cite{journals/tkde/ShangLLF16}; \textit{personalised medicine} provides specific treatment to a small group of patients clustered by a disease taxonomy \cite{giovannoni2017personalized}. Also, taxonomic joins can be helpful for enhancing the quality of similarity matrices used in various recommender systems \cite{conf/sigir/ZhangL0ZLM14}.

In this article, we tackle the taxonomic joining problem by adopting the \textit{filtering-and-verification} framework, which works by first (i) removing record pairs which are impossible to be similar, then (ii) verifying the real similarity of survived pairs. Since verification is expensive, we introduce a novel filtering technique which efficiently removes unfeasible pairs whose number of similar nodes do not reach a given \textit{overlap constraint}. We observed that such constraint vitally affects filtering quality and thus joining time. In this paper, we propose an estimator to predict the running time of a given constraint by performing test-drives on small samples and then scaling the result accordingly. To suggest the best constraint for minimising the joining time, as our technical contribution, we propose a Monte Carlo simulation process which gives accurate suggestions without a predetermined sample size. Experiments show that, remarkably, our suggestion achieves higher than 90\% accuracy, by using only a few (e.g., 100) samples from nearly 3 million records, and occupies approximately 1\% of total joining time. The superior performance of our solution over the state-of-art approach \cite{journals/tkde/ShangLLF16} further motives its application in practice.


\section{Preliminaries}
\label{sec:preliminaries}


\noindent\textbf{Similarity measures.}
Let $S:\{s_1,\cdots,s_i\}$ and $T:\{t_1,\cdots,t_j\}$ be two sets of nodes from a hierarchical taxonomy. Let $s \in S$ and $t \in T$ be two nodes, and let $|s|$ ($|t|$) denotes the depth of node $s$ ($t$). Then, their similarity can be measured based on the depth of the lowest common ancestor (LCA):
\begin{equation}
    \label{eq:ts_ic}
    TS(s,t) = \frac{|LCA(s,t)|}{\max(|s|,|t|)}
\end{equation}

With Equation \ref{eq:ts_ic}, the similarity between two sets $S$ and $T$ can then be obtained by averaging the maximum sum of all $TS$'s of \textit{distinct} node pairs, where $|S|$ ($|T|$) is the number of nodes in set $S$ ($T$), $I_{pq}$ is an indicator variable (i) controlling whether to select the edge $(s_p,t_q)$, and (ii) ensures any of $s_p$ or $t_q$ is used at most once:
\begin{equation}
    \label{eq:gts_ic}
    \begin{split}
        &GTS(S,T) = \frac{W(S,T)}{\max(|S|,|T|)} = \frac{\max\textstyle\sum_p\sum_q I_{pq} TS(s_p,t_q)}{\max(|S|,|T|)} \\
        &where~p\in[1,i],q\in[1,j],I_{pq}=0~or~1,\textstyle\sum_p I_{pq} \leq 1,and~\textstyle\sum_q I_{pq} \leq 1 \\
    \end{split}
    \raisetag{30pt}
\end{equation}


Solving for the value of $W$ in Equation \ref{eq:gts_ic} requires to find the maximum weight matching in a bipartite graph, which can be categorised as an \textit{assignment problem}. \textit{Hungarian algorithm} \cite{journals/JSIAM/Munkres57} is so far the best solution which runs in a polynomial $\mathcal{O}(n^3)$ time.

\begin{example}
    \label{exp:gts}
    Take two strings in Figure \ref{fig:toytaxonomyexample} as an example. Since the three most-similar node pairs are (``coffeehouse'', ``bar''), (``latte'', ``espresso''), and (``Turin'', ``Via Nizza''), the $GTS$ similarity between two strings becomes 0.717 ($=(0.75+0.8+0.6)/3$). Note that the distinctness forbid any node from being selected more than once, e.g., selecting both (``latte'', ``espresso'') and (``latte'', ``Turin'') are not allowed.
\end{example}

\noindent\textbf{Problem definition.}
We define our research problem as follows:
\begin{problem}
    \label{prob:main_prob}
    Let $\mathcal{S}$ and $\mathcal{T}$ be two collections of sets $\mathcal{S}:\{S_1,\cdots,S_m\}$, $\mathcal{T}:\{T_1,\cdots,T_n\} $, where each set contains multiple nodes, i.e., $S\in\mathcal{S}:\{s_1,\cdots,s_i\} $, $T\in\mathcal{T}:\{t_1,\cdots,t_j\}$. Given a $GTS$ similarity measure and a similarity threshold $\theta$, find all pairs of sets in forms of $(S,T)\in \mathcal{S} \times \mathcal{T}$ such that each $GTS(S,T) \geq \theta$.
\end{problem}

It is not trivial to solve Problem \ref{prob:main_prob} efficiently. Recall Example \ref{exp:gts}. To apply the Hungarian algorithm, we first need to fill a $3\times 3$ matrix by $9$ TS calculations, not to mention a longer string which may have hundreds of words resulting in a massive amount of calculations. Large datasets exacerbate the situation as processing every string pair leads to interminable running time. Hence, it is crucial to have an efficient solution which avoids running Hungarian algorithm whenever possible to speed up the joining process.

\section{Adaptive overlap joining}
\label{sec:multi_joins}

We now present our novel joining algorithm which include three stages, namely (i) \textit{inverted lists construction} (Line 2 in Alg. \ref{alg:ss_join}), where each set is being indexed for faster overlap-finding; (ii) \textit{filtering} (Lines 3 - 14), where we try to purge unfeasible set pairs which have not enough ($\geq\tau$) similar nodes; and (iii) \textit{verification} (Lines 15 - 16) where we perform actual $GTS$ calculation on survived pairs. Since Stage 2 is a key step to speed up the whole processing, we will focus on developing an effective optimisation strategy.

Based on the definition of $GTS$, our filtering technique states that two similar sets must have at least $\tau$ pairs of similar nodes, where $\tau$ is the overlap constraint:

\begin{lemma}[AP-Filter]
    \label{lem:ss_filter}
    Given two sets $S$ and $T$, and without loss of generality by assuming $|S|<|T|$. If $GTS(S,T) \geq \theta$, then there are at least $\tau$ distinct similar pairs of nodes $(s_i \in S, t_j \in T)$ such that each of them satisfies $TS(s_i, t_j) \geq \varphi$, where $\varphi=\frac{\theta |T| - \tau +1}{|S|- \tau +1}$.
\end{lemma}

According to Lemma \ref{lem:ss_filter}, for each pair of sets, we need to find the number of distinct node pairs whose $TS\geq\varphi$. Since $TS$ depends on $|LCA|$, the problem can be converted to overlap finding problem, which is to find the common ancestors between pairs of nodes within two sets. An efficient way to find such ancestors is to index them (as keys, corresponding nodes as values) by \textit{inverted lists}, and joining these lists afterwards to obtain overlapped keys and thus sets which contains nodes having common ancestors. Furthermore, Lemma \ref{lem:ss_filter} can be relaxed to the follows, so that we can build an inverted list independently for each of $\mathcal{S}$ and $\mathcal{T}$:

\begin{corollary}
    \label{cor:tas_phi_threshold}
    Two similar sets, $S$ and $T$, must have at least $\tau$ distinct $(s_i, t_j)$'s such that each $TS(s_i, t_j) \geq \varphi$, where $\varphi = \frac{\theta |S| - \tau +1}{|S|-\tau+1}$.
\end{corollary}

Corollary \ref{cor:tas_phi_threshold} states that a node $s \in S$ only needs to put its ancestors deeper than $\varphi |s|$ into the inverted list since others are not able to achieve $\varphi$ similarity. This insight reduces the index size and thus accelerate filtering.

    

\begin{algorithm}[b]
    \caption{\textsc{AP-Join}: Set joining with \textsc{AP-Filter}}
    \label{alg:ss_join}
    \KwIn{two collections of sets $\mathcal{S}$ and $\mathcal{T}$, a similarity threshold $\theta$ and an positive integer $\tau$}
    \KwOut{$R$: $\{(S,T) \in \mathcal{S} \times \mathcal{T} | GTS(S,T) \geq \theta\}$}
    
    $\mathcal{P}$ $\gets$ $\varnothing$, $\mathcal{C}$ $\gets$ $\varnothing$, $\mathcal{R}$ $\gets$ $\varnothing$\\
    $\mathcal{L}_{\mathcal{S}},\mathcal{L}_{\mathcal{T}}$ $\gets$ inverted lists built from $\mathcal{S}$ and $\mathcal{T}$, according to Cor. \ref{cor:tas_phi_threshold} \\
    $\mathcal{G}$ $\gets$ overlapped keys (common ancestors) between $\mathcal{L}_{\mathcal{S}}$ and $\mathcal{L}_{\mathcal{T}}$ \\
    \ForEach{$g \in \mathcal{G}$}{
        $\ell_{\mathcal{S}}, \ell_{\mathcal{T}} \gets$ lists indexed by $g$ in $\mathcal{L}_{\mathcal{S}}$ and $\mathcal{L}_{\mathcal{T}}$ \\
        \ForEach{$(S,T) \in (\ell_{\mathcal{S}} \times \ell_{\mathcal{T}})$}{
            \If{$\min(|S|,|T|) \leq \theta \cdot \max(|S|,|T|$)}{
                \textbf{continue} \tcp*{length filtering}
            }
            $s_g,t_g \gets$ nodes producing ancestor $g$ in $S$ and $T$\\
            \lIf{$s_g$ and $t_g$ have not been found similar with other nodes from $T$ or $S$}{
                $\mathcal{P} \gets \mathcal{P} \cup \{(S,T)\}$
            }
        }
    }
    \ForEach(\tcp*[f]{find pairs with $\tau$ overlaps}){$(S,T) \in \mathcal{P}$}{
        \If{$(S,T)$ appears at least $\tau$ times in $\mathcal{P}$}{
            $\mathcal{C}$ $\gets$ $\mathcal{C} \cup \{(S,T)\}$\\
        }
    }
    \ForEach(\tcp*[f]{verification}){$(S,T) \in \mathcal{C}$}{
        \lIf{$GTS(S,T) \geq \theta$}{
            $\mathcal{R}$ $\gets$ $\mathcal{R} \cup \{(S,T)\}$
        }
    }
    \Return{$\mathcal{R}$}
\end{algorithm}

The pseudo code of our joining technique is presented in Algorithm \ref{alg:ss_join}. Specifically, Line 7 employs a \textit{length filtering} technique to remove pairs having a disparate number of nodes, and Line 10 ensures the distinctness of each node so that a node $s$ is counted one time even if it is similar to multiple $t$'s from another set $T$.

\noindent\textbf{Effect of overlapping constraint $\mathbf{\tau}$.}
The value of parameter $\tau$ influences the joining time. Intuitively, as $\tau$ increases, the sizes of inverted lists will grow, indicating that the \textit{filtering} will be slower. In the meantime, fewer pairs satisfies the increased overlap constraint, leading to the faster \textit{verification} phase. Perceiving the opposite trends, we conduct an empirical experiment as presented in Figure \ref{fig:exp_tau}, in which we confirm the existence of optimum that minimises the overall running time. Now a natural question arises: \textit{how to find such optimal $\tau$ correctly and efficiently?} We tackle the question and present our answer in the next section.

\begin{figure}
    \hspace*{-13mm}
    \begin{tikzpicture}
	\pgfplotstableread{
		y	tau1	tau2	tau3	tau4	tau5
		0.6	4.41	36.31	43.76	53.59	68.28
		0.7	1.46	18.56	19.04	19.63	25.55
		0.8	0.38	5.59	6.93	6.87	7.78
	}\dataP
	\pgfplotstableread{
		y	tau1	tau2	tau3	tau4	tau5
		0.6	2.53	0.85	0.27	0.14	0.07
		0.7	0.89	0.55	0.12	0.03	0.02
		0.8	0.35	0.23	0.07	0.02	0.01
	}\dataC
	\pgfplotstableread{
		y	tau1	tau2	tau3	tau4	tau5
		0.6	74.46	42.80	25.88	27.45	30.93
		0.7	26.97	21.95	12.69	10.17	12.99
		0.8	10.03	10.82	5.93	4.65	5.04
	}\dataT
	\scriptsize
	\begin{groupplot}[
		group style={group name=plots,group size=3 by 1, horizontal sep=3.5em},
		width=3.5cm,
		height=2.8cm,
		ybar=0pt,
		xmin=0.6,xmax=0.8,
		xtick={0.6,0.7,0.8},
		enlarge x limits=0.3,
		enlarge y limits=0,
		xtick align=inside,
		area legend,
		legend cell align=left,
		legend columns=5,
		legend image code/.code={\draw [#1] (0cm,-0.1cm) rectangle (0.2cm,0.1cm);},
		legend style={at={(1,1)},anchor=south east,draw=none,fill=none},
	]
		\nextgroupplot[bar width=2.5pt,ylabel={Pairs ($\times10^6$)},ymin=0,ymax=60]
			\addplot+ table[y = tau1] {\dataP};
			\addplot+ table[y = tau2] {\dataP};
			\addplot+ table[y = tau3] {\dataP};
			\addplot+ table[y = tau4] {\dataP};
			\addplot+ table[y = tau5] {\dataP};

		\nextgroupplot[bar width=2.5pt,ylabel={Candidates ($\times10^6$)},xlabel={Joining similarity threshold $\theta$},ymin=0,ymax=3]
			\addplot+ table[y = tau1] {\dataC};
			\addplot+ table[y = tau2] {\dataC};
			\addplot+ table[y = tau3] {\dataC};
			\addplot+ table[y = tau4] {\dataC};
			\addplot+ table[y = tau5] {\dataC};

		\nextgroupplot[bar width=2.5pt,ylabel={Time (s)},ymin=0,ymax=50]
			\addplot+ table[y = tau1] {\dataT};
			\addplot+ table[y = tau2] {\dataT};
			\addplot+ table[y = tau3] {\dataT};
			\addplot+ table[y = tau4] {\dataT};
			\addplot+ table[y = tau5] {\dataT};

			\legend{$\tau=1$,$\tau=2$,$\tau=3$,$\tau=4$,$\tau=5$}
	\end{groupplot}

	\node [text width=6cm,align=center,anchor=north] at ([yshift=-4mm]plots c1r1.south)
	{\subcaption{Pairs in inverted lists}};
	\node [text width=6cm,align=center,anchor=north] at ([yshift=-4mm]plots c2r1.south)
	{\subcaption{Candidates}};
	\node [text width=6cm,align=center,anchor=north] at ([yshift=-4mm]plots c3r1.south)
	{\subcaption{Running time}};
\end{tikzpicture}
    \vspace*{-8mm}
    \caption{Overlap constraint $\mathbf{\tau}$'s affecting joining performance ($\mathbf{10K \times 10K}$ subsets of OHSUMED).}
    \vspace*{-5mm}
    \label{fig:exp_tau}
\end{figure}

\section{Parameter recommendation}
\label{sec:adaptive_filtering}

This section aims to give an accurate recommendation for $\tau$ which minimises the total joining time. It is backed by a cost model and a sampling-based estimator.

\noindent\textbf{Cost model.}
The joining time cost can be modelled as follows:
\begin{equation}
    \label{eq:cost_func}
        C_\tau=C_{F_\tau}+C_{V_\tau}= t_F \cdot F_\tau + t_V \cdot V_\tau
\end{equation}
\noindent where the total cost $C_\tau$ is the sum of filtering ($C_{F_\tau}$), and verification cost ($C_{V_\tau}$). Each of them is obtained by multiplying corresponding number of processed pairs ($F_\tau$ or $V_\tau$) by the average time to process one pair in each stage ($t_F$ or $t_V$).

\noindent\textbf{Bernoulli estimator.}
It is certainly unfeasible to run the joining algorithm on full datasets to get its cost. Instead, we can \textit{estimate} the number of pairs in each stage ($\hat{F}_\tau$ and $\hat{V}_\tau$) and hence the total cost $\hat{C}_\tau$ by using the \textit{independent Bernoulli sampling}, where each set in input dataset $\mathcal{S}$ ($\mathcal{T}$) has probability $p_s$ ($p_t$) for being in the sample. Therefore, a set pair $(S,T)$ being processed during a \textit{real} filtering (or verification) stage has probability $p_s p_t$ for being counted into $F_\tau$ (or $V_\tau$), i.e., when once both $S$ and $T$ exist in the sample. Hence, we get an unbiased estimator of $F_\tau$:
\begin{equation}
    E[F_\tau'] = F_\tau \cdot p_s p_t \Rightarrow \hat{F}_\tau= \frac{F_\tau'}{p_s p_t}.\text{~Similarly,~}\hat{V}_\tau= \frac{V_\tau'}{p_s p_t}
    \label{eq:e_j_hat}
\end{equation}
Plugging $\hat{F}_\tau$ and $\hat{V}_\tau$ into Equation \ref{eq:cost_func} to obtain estimated cost $\hat{C}_\tau$.

\noindent\textbf{Iterative suggestion refinement.}
Equation \ref{eq:e_j_hat} is a static estimation strategy where $p_s$ and $p_t$ are determined beforehand, usually by trial-and-error. To erase the requirement of this foreknowledge, we propose an iterative method based on Monte Carlo simulation, which refines the suggestion from multiple iterations until the smallest $\hat{C}_\tau$ is identified with high confidence.

Multiple iterations give a series of estimations. Since all of them are i.i.d., the Central Limit Theorem (CLT) holds such that their mean $\hat{\mu}_{F_\tau}$ converges to a normal distribution when iteration goes on (the same for $\hat{\mu}_{V_\tau}$; $Var[\cdot]$ denotes the population variance):
\begin{equation}
    \hat{\mu}_{F_\tau} \sim \mathcal{N}\big(E[\hat{F}_\tau],Var[\hat{F}_\tau]/n\big)~when~n \rightarrow \infty
\end{equation}

CLT also allude that the mean and variance of underlying distribution, $E[\hat{F}_\tau]$ and $Var[\hat{F_\tau]}/n$, can be estimated by the sample mean and variance, $\hat{\mu}_{F_\tau}$ and $\hat{\sigma}_{F_\tau}^2$. Both estimators are unbiased. Thus, we can calculate $\hat{\mu}_{F_\tau}$ and $\hat{\sigma}_{F_\tau}^2$ by a recursive formula (e.g., \cite{journals/cam/finch2009}), and use them to estimate $\hat{\mu}_{F_\tau}$ (the same for $\hat{\mu}_{V_\tau}$). Since both $\hat{F}_\tau$ and $\hat{V}_\tau$ converge to normal distributions, the estimated total cost $\hat{C}_\tau$ also converges, and can be modelled by a \textit{Student's $t$ distribution}. The confidence interval (CI) of $C_\tau$ can be constructed consequently.


\noindent\textbf{Stopping criterion.}
Given a universe of $\tau$'s, we can safely terminate the refinement procedure once the overlapped CI's between the best (which gives the least $\hat{\mu}_{C_\tau}$) and other $\tau$'s are small enough. In other words, the refinement stops when \textit{the additional cost due to an inaccurate estimation (i.e., the sum of overlapped CI's) is less than that for one more iteration}:
\begin{lemma}[Stopping Criterion]
    \label{lem:stopping_criterion}
    Let $\mathbb{U}_\tau$ be the universe of $\tau$'s, and let $\tau_1\in\mathbb{U}_\tau$ denote the $\tau$ leading to the minimal estimated cost, i.e., $\tau_1=\arg\min_{\tau}\left(\hat{\mu}_{C_\tau}\right)$. The refinement process terminates when $\sum_{\tau_2}(U_{C_{\tau_1}} - L_{C_{\tau_2}}) < { t_F \cdot \textstyle \sum}_\tau F_\tau'$ holds for all $\tau_2\in\mathbb{U}_\tau,\tau_2\neq\tau_1$.
\end{lemma}

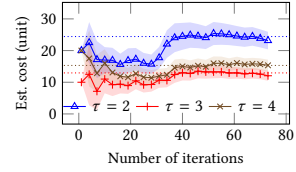
\begin{figure}[t]
    \centering
    \begin{tikzpicture}
	\pgfplotstableread{
		x	tau1	tau2	tau3	tau4	tau5	err1	err2	err3	err4	err5
		1	110000	20000	10000	20000	370000	0	0	0	0	0
		2	155000	20000	5000	15000	290000	46620	0	5180	5180	82880
		3	153333	26667	16667	23333	273333	26971	6907	12451	9137	50871
		4	127500	22500	12500	17500	260000	32863	6518	9806	8847	38532
		5	124000	18000	10000	14000	208000	25713	6872	8025	7753	61588
		6	138333	20000	8333	15000	256667	25715	5981	6776	6414	71209
		7	121429	17143	7143	12857	248571	27912	5858	5858	5858	60764
		8	125000	17500	8750	15000	251250	24454	5087	5339	5538	52696
		9	133333	16667	10000	15556	240000	23230	4568	4884	4918	47913
		10	127000	17000	11000	16000	228000	21789	4101	4489	4422	44622
		11	115455	15455	10000	14545	214545	23054	4040	4191	4275	42701
		12	135000	15000	10000	13333	210000	29205	3718	3826	4099	39264
		13	133077	16923	9231	13077	210000	26939	3958	3608	3780	36117
		14	132857	15714	9286	12143	217857	24941	3872	3341	3631	34415
		15	128000	16000	9333	12000	222667	23758	3617	3111	3384	32424
		16	123125	15625	9375	11875	218750	22790	3406	2910	3168	30600
		17	125294	15294	8824	11176	223529	21525	3217	2793	3062	29167
		18	125000	17778	9444	12222	224444	20297	3978	2710	3084	27515
		19	118947	16842	8947	11579	246842	20197	3885	2615	2992	34868
		20	118000	18000	10500	13000	241000	19186	3876	2957	3198	33628
		21	114762	17619	10476	12857	233810	18555	3708	2812	3045	32843
		22	112273	17273	10455	12727	232273	17878	3554	2682	2907	31355
		23	119565	16522	10000	12609	239130	18679	3484	2605	2780	30791
		24	121667	16667	9583	12083	237917	18016	3339	2531	2717	29507
		25	119600	16000	9200	11600	238400	17413	3276	2460	2653	28307
		26	117692	15769	9231	11538	231923	16846	3157	2364	2550	28012
		27	122593	15185	8889	11111	229259	16986	3097	2302	2494	27095
		28	126071	15714	9286	11429	238929	16760	3034	2256	2425	27965
		29	126207	16552	10345	12069	237586	16173	3054	2438	2432	27020
		30	124000	16000	10000	11667	238000	15791	3005	2382	2387	26107
		31	129677	17742	10645	11935	240000	16366	3421	2399	2325	25336
		32	130312	17813	10625	11875	242813	15860	3313	2323	2252	24704
		33	131212	20000	10606	12121	248182	15400	3931	2252	2198	24581
		34	138235	22059	10588	12353	257353	16618	4369	2184	2145	25670
		35	141429	22000	10571	12571	254000	16472	4243	2121	2095	25167
		36	154722	24167	12222	14167	261389	21117	4695	2678	2623	25628
		37	154865	24054	12432	14595	260541	20539	4568	2614	2589	24941
		38	152105	24211	12632	14474	260526	20194	4449	2553	2523	24276
		39	149231	24103	12821	14872	260769	19894	4335	2494	2492	23646
		40	149500	24500	13000	15000	259500	19392	4245	2438	2432	23085
		41	147317	24146	12927	14878	254878	19048	4156	2379	2376	23019
		42	147857	24286	12857	14762	259048	18597	4059	2323	2322	22875
		43	146047	24884	12791	14651	260000	18256	4011	2269	2270	22359
		44	146136	24318	12500	14318	268182	17837	3963	2237	2244	23432
		45	148889	24222	12889	14444	266222	17668	3875	2224	2198	22995
		46	150000	24565	13478	15217	273696	17318	3806	2259	2294	23785
		47	148298	24043	13191	14894	270000	17037	3764	2230	2269	23586
		48	149375	24167	13125	14792	272917	16715	3687	2185	2224	23286
		49	148776	24082	13265	14898	270204	16382	3612	2144	2181	22979
		50	150400	24200	13400	15600	275400	16139	3541	2106	2257	23149
		51	152745	25490	13333	15882	279216	16005	3719	2065	2232	23032
		52	155192	25385	13269	15962	280577	15898	3649	2026	2190	22629
		53	153962	25283	13396	16226	284717	15647	3581	1992	2166	22609
		54	152407	25185	13519	16296	281667	15439	3515	1959	2127	22410
		55	152909	25273	13273	16000	283273	15164	3452	1940	2110	22061
		56	151250	25179	13393	16071	282500	14990	3391	1909	2073	21679
		57	150175	24912	13158	15789	281930	14767	3342	1891	2057	21303
		58	149655	25000	12931	15517	281207	14520	3286	1873	2041	20946
		59	149153	24576	12881	15424	280000	14281	3259	1841	2009	20626
		60	148667	24500	12667	15833	282667	14050	3205	1824	2020	20467
		61	149836	24590	12951	16066	286721	13871	3154	1818	2001	20562
		62	150806	24839	12903	15806	289677	13682	3113	1789	1987	20458
		63	149524	24603	12857	15714	294127	13529	3073	1761	1957	20652
		64	149063	24219	12656	15625	292344	13324	3051	1746	1929	20410
		65	147538	24000	12615	15538	289846	13212	3012	1719	1901	20260
		66	150606	24394	12576	15455	293030	13393	2994	1693	1874	20221
		67	151642	24478	12836	15821	292090	13235	2950	1690	1884	19941
		68	151765	24412	12794	16029	290588	13040	2907	1665	1869	19707
		69	152754	24203	12754	15942	289565	12890	2873	1641	1844	19448
		70	151000	23857	12571	15714	286143	12834	2854	1629	1833	19494
		71	150141	23521	12394	15493	284507	12683	2835	1616	1821	19292
		72	151111	23472	12222	15556	285972	12546	2796	1603	1797	19082
		73	149589	23151	12055	15342	282740	12473	2777	1591	1786	19115
		74	148784	22838	11892	15270	283243	12332	2759	1578	1763	18862
	}\tabledata
	
	\scriptsize
	\begin{axis}[
		width=4.6cm,
		height=3.2cm,
		enlarge x limits=0.1,
		enlarge y limits=0.1,
		scaled y ticks=base 10:-3,
		xtick align=inside,
		each nth point=3,
		xmin=1,xmax=74,
		ymin=0,ymax=32000,
		scaled y ticks=base 10:-3,
		ytick scale label code/.code={},
		every y tick scale label/.style={},
		xlabel={Number of iterations},
		mark size=1.5,
		ylabel={Est. cost (unit)},
		scatter/classes={
			1={mark=triangle,blue},
			2={mark=+,red},
			3={mark=x,brown!60!black},
			4={mark=star,black},
			5={mark=diamond,violet}
		},
        legend columns=3,
		legend cell align=left,
		legend style={at={(1,0)},anchor=south east,draw=none,fill=none},
		legend image code/.code={
			\draw[mark repeat=2,mark phase=2]
				plot coordinates {
				(0cm,0cm)
				(0.15cm,0cm)        
				(0.3cm,0cm)         
				};%
		},
	]
		\errorband{\tabledata}{x}{tau2}{err2}{blue}{0.15}
		\addplot+[scatter, scatter src=1]
		table[x = x,y = tau2] {\tabledata};
		\addlegendentry{$\tau=2$}
		
		\errorband{\tabledata}{x}{tau3}{err3}{red}{0.15}
		\addplot+[scatter, scatter src=2]
		table[x = x,y = tau3] {\tabledata};
		\addlegendentry{$\tau=3$}
		
		\errorband{\tabledata}{x}{tau4}{err4}{brown}{0.15}
		\addplot+[scatter, scatter src=3]
		table[x = x,y = tau4] {\tabledata};
		\addlegendentry{$\tau=4$}
		
		\draw [densely dotted,blue] (-100,24436) -- (900,24436);
		\draw [densely dotted,red] (-100,12949) -- (900,12949);
		\draw [densely dotted,brown!60!black] (-100,15298) -- (900,15298);
	\end{axis}
\end{tikzpicture}
    \vspace*{-4mm}
    \caption{Illustration of the convergent of the means (solid lines) and CI's (shaded areas) to the real values (dotted lines). The cost when $\mathbf{\tau=1}$ is too large thus have been eliminated.}
    \vspace*{-5mm}
    \label{fig:tau_selection_example}
\end{figure}

\begin{algorithm}[b]
    \caption{Cost-bounded suggestion refinement}
    \label{alg:tau_estimation}
    \KwIn{$k$ samples from each dataset: $\big\{\mathcal{S}_1', \cdots, \mathcal{S}_k'\big\}$ and $\big\{\mathcal{T}_1', \cdots, \mathcal{T}_k'\big\}$, a positive integer $n^*$, a Student's $t$ quantile $t_*$ corresponding to a specific confidence level, and a universe of $\tau$'s $\mathbb{U}_\tau$}
    \KwOut{$\tau$ which has the minimal estimated cost}
    
    $n$ $\gets$ 1 \tcp*{select first samples}
    \Repeat{$n \geq n^*$ and the condition in Lemma \ref{lem:stopping_criterion} is fulfilled}{
        \ForEach{$\tau\in\mathbb{U}_\tau$}{
             Run the filtering stage of \textsc{AP-Join} on samples $\mathcal{S}_n'$ and $\mathcal{T}_n'$ \\
            Compute the confidence interval of total cost $CI(C_\tau)$ \\
        }
        $n$ $\gets$ $n+1$ \tcp*{select next samples}
    }
    \Return{$\tau_1$ in Lemma \ref{lem:stopping_criterion}}
\end{algorithm}

We present our iterative refinement procedure in Algorithm \ref{alg:tau_estimation}. Given multiple independent samples form $\mathcal{S}$ and $\mathcal{T}$, it runs the filtering stage of \textsc{AP-Join} for every $\tau$, obtaining $F_\tau'$ and $V_\tau'$, then estimates the mean and variance of $\hat{C}_\tau$. The procedure terminates when the best $\tau$ is found with a predefined confidence level. The refinement runs at least $n^*$ iterations to discard the effect of instability in the early stage, known as the \textit{burn-in} period.

\section{Experimental analysis}
\label{sec:experiments}

\begin{table}[t]
    \caption{Characteristics of taxonomies and datasets.}
    \label{tab:datasets}
    \scriptsize
    \vspace*{-4mm}
    \addtolength{\tabcolsep}{-3pt} 
\setlength\extrarowheight{-2pt}
\begin{tabular}{@{}cccccccc@{}}
\toprule
\multicolumn{4}{c}{Taxonomy trees (Height in min/avg/max)}                         &  & \multicolumn{3}{c}{String datasets (\# nodes in min/avg/max)}                         \\[-1px] \cmidrule(){1-4} \cmidrule(){6-8} 
Name            & \# of nodes & Height       & Avg. fanout &  & Name          & \# of records & \# of nodes \\ \midrule
MeSH            & 57,840      & 1 / 5.1 / 12 & 157         &  & OHSUMED       & 293,294       & 5 / 8.4 / 26                \\
Wiki categories & 1,212,943   & 1 / 6.2 / 26 & 32,300      &  & Wiki articles & 3,512,954     & 5 / 8.2 / 277               \\ \bottomrule

\end{tabular}
\addtolength{\tabcolsep}{3pt} 
    \vspace*{-3mm}
\end{table}

We implemented all algorithms in Java 8, and run the code on a quad-core Xeon 2.53GHz node with 32GB RAM. We use two taxonomies, \textit{Wiki categories}\footnote{\url{http://wiki.dbpedia.org/services-resources/documentation/datasets}} and \textit{MeSH terms}\footnote{\url{https://www.nlm.nih.gov/mesh} and \url{http://trec.nist.gov/data/t9_filtering.html}}, and two string datasets, \textit{Wiki articles} and \textit{OHSUMED articles}. Each string is mapped to a set of taxonomic nodes according to its categories (Wiki articles) or keywords (OHSUMED). Table \ref{tab:datasets} describes these datasets.



\noindent\textbf{Performance vs the state-of-art.}
We obtained the source code from the authors of the state-of-art approach \textsc{K-Join} \cite{journals/tkde/ShangLLF16}, rewrote their C++ code using Java, and extended the algorithm to perform R-S join following their instructions. After that, we ran both \textsc{AP-Join} and \textsc{K-Join} to compare their performance and present Table \ref{tab:filtering_power}. It shows that our approach outperforms the state-of-art on both datasets, especially OHSUMED, with a large margin. Specifically, given the threshold 0.6, \textsc{K-Join} uses up 32GB of RAM for storing all pairs and eventually crashed. \textsc{AP-Join}, on the contrary, has fewer candidates and successfully finishes the joining.

\begin{table}
    \caption{Performance vs the state-of-art w.r.t. $\mathbf{\theta}$.}
    \label{tab:filtering_power}
    \scriptsize
    \vspace*{-4mm}
    \addtolength{\tabcolsep}{-2.5pt} 
\setlength\extrarowheight{-2pt}
\begin{tabular}{@{}ccccccccccccc@{}}
	\toprule
	\multirow{2}{*}{Dataset} & \multirow{2}{*}{Algorithm} & \multicolumn{3}{c}{\# Pairs ($10^8$)} &  & \multicolumn{3}{c}{\# Candidates ($10^6$)} &  & \multicolumn{3}{c}{Running time (min)} \\ \cmidrule(lr){3-5} \cmidrule(lr){7-9} \cmidrule(l){11-13} 
	 &  & 0.6 & 0.7 & 0.8 &  & 0.6 & 0.7 & 0.8 &  & 0.6 & 0.7 & 0.8 \\ [-1px] \midrule
	\multirow{2}{*}{\begin{tabular}[c]{@{}c@{}}Wiki articles\\ (50K$\times$50K)\end{tabular}} & \textsc{AP-Join} & \textbf{0.42} & \textbf{0.08} & \textbf{0.01} &  & \textbf{11.52} & \textbf{3.04} & \textbf{0.27} &  & \textbf{10.03} & \textbf{2.64} & \textbf{0.55} \\
	 & \textsc{K-Join} & 0.87 & 0.25 & 0.07 &  & 28.42 & 8.35 & 1.98 &  & 22.28 & 6.65 & 1.74 \\ \midrule
	\multirow{2}{*}{\begin{tabular}[c]{@{}c@{}}OHSUMED\\ (50K$\times$50K)\end{tabular}} & \textsc{AP-Join} & \textbf{1.08} & 4.97 & 1.72 &  & \textbf{63.43} & \textbf{0.64} & \textbf{0.26} &  & \textbf{41.81} & \textbf{4.44} & \textbf{1.67} \\
	 & \textsc{K-Join} & - & \textbf{2.13} & \textbf{0.86} &  & - & 115.58 & 38.42 &  & - & 80.01 & 25.33 \\ \bottomrule
	\end{tabular}
\addtolength{\tabcolsep}{2.5pt} 
\setlength\extrarowheight{0pt}

    \vspace*{-3mm}
\end{table}

\begin{table}
    \caption{Suggestion accuracy and speed (100 samples).}
    \label{tab:mc_correctness}
    \scriptsize
    \vspace*{-4mm}
    \addtolength{\tabcolsep}{-2.5pt} 
\setlength\extrarowheight{-2pt}

\begin{tabular}{@{}ccccccccccc@{}}
    \toprule
    \multirow{2}{*}{Dataset} & \multirow{2}{*}{} & \multicolumn{4}{c}{Accuracy from 128 runs varies $\theta$} &  & \multicolumn{4}{c}{Estimation time varies $\theta$ (s)} \\ \cmidrule(l){3-6} \cmidrule(){8-11} 
                             &                   & 0.6            & 0.7           & 0.8           & 0.9         &  & 0.6          & 0.7          & 0.8         & 0.9         \\[-1px] \midrule
    Wiki articles            &                   & 92.03\%        & 100\%         & 100\%         & 100\%       &  & 2.58         & 1.69         & 1.74        & 1.51        \\
    OHSUMED                  &                   & 96.09\%        & 99.22\%       & 90.63\%       & 100\%       &  & 4.63         & 1.10         & 2.04        & 1.42        \\ \bottomrule
\end{tabular}
\setlength\extrarowheight{0pt}
\addtolength{\tabcolsep}{2.5pt} 

    \vspace*{-3mm}
\end{table}

\begin{figure}
    \centering
    \hspace*{-5mm}
    \begin{tikzpicture}
	\pgfplotstableread{
		size	0.8	0.9
		10	0.06	0.01
		30	0.49	0.03
		50	1.45	0.09
		70	2.71	0.15
		90	4.91	0.26
		110	7.31	0.39
		130	9.62	0.55
		150	12.98	0.74
	}\dataMESH
	\pgfplotstableread{
		size	0.8	0.9
		0.5	25.33	1.55
		0.7	47.16	3.45
		0.9	79.87	5.03
		1.1	110.00	7.01
		1.3	110.00	9.59
		1.5	110.00	12.69
	}\dataWIKI
	\scriptsize
	\begin{groupplot}[
		group style={group name=plots,group size=2 by 1, horizontal sep=5em},
		width=4cm,
		height=2.8cm,
		enlarge x limits=0.1,
		enlarge y limits=0.1,
		xtick align=inside,
		scatter/classes={
			1={mark=triangle,blue},
			2={mark=+,red},
			3={mark=x,brown!60!black},
			4={mark=star,black},
			5={mark=diamond,violet}
		},
        legend columns=2,
		legend cell align=left,
		legend style={at={(1,1)},anchor=south east,draw=none,fill=none},
		legend image code/.code={
			\draw[mark repeat=2,mark phase=2]
				plot coordinates {
				(0cm,0cm)
				(0.15cm,0cm)        
				(0.3cm,0cm)         
				};%
		},
	]
		\nextgroupplot[xlabel={Size of both datasets (K)},ylabel={Time (min)},xtick={10,30,50,70,90,110,150},ymin=0,ymax=15]
			\addplot+[scatter, scatter src=1]
			table[y = 0.8] {\dataMESH};
			\addplot+[scatter, scatter src=2]
			table[y = 0.9] {\dataMESH};
		
		\nextgroupplot[xlabel={Size of both datasets (M)},ylabel={Time (min)},xtick={0.5,0.7,0.9,1.1,1.3,1.5},ymin=0,ymax=80]
		\addplot+[scatter, scatter src=1]
		table[y = 0.8] {\dataWIKI};
		\addplot+[scatter, scatter src=2]
		table[y = 0.9] {\dataWIKI};
		\legend{$\theta=0.8$,$\theta=0.9$}
	\end{groupplot}

	\node [text width=6cm,align=center,anchor=north] at ([yshift=-4mm]plots c1r1.south)
	{\subcaption{OHSUMED}};
	\node [text width=6cm,align=center,anchor=north] at ([yshift=-4mm]plots c2r1.south)
	{\subcaption{Wiki articles}};
\end{tikzpicture}
    \vspace*{-8mm}
    \caption{Scalability experiments on (a) mid-sized and (b) large-sized datasets. Two joining datasets have the same size.}
    \vspace*{-4mm}
    \label{fig:scalability}
\end{figure}
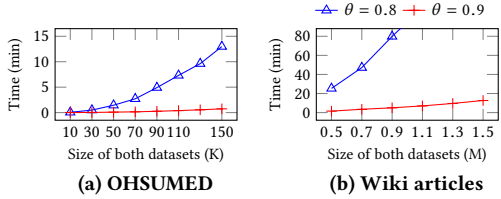

\noindent\textbf{Scalability.}
We randomly sample our datasets into different subsets so that the largest one contains roughly half of total records. Then, we run our algorithm and present the result in Figure \ref{fig:scalability}. The result shows that the joining time increases linearly with data size. Besides, a larger $\theta$ leads to a faster joining, because a high similarity threshold leads to a high $\varphi$, which reduces sizes of inverted lists, the number of candidates, and ultimately the total joining time.

\noindent\textbf{Parameter suggestion.}
The final experiment is the iterative procedure which suggests the key parameter $\tau$. We set $n^*=10$, $t_*=1.036$ (70\% confidence level) following common practices, and the sample size be 100, a tiny fraction comparing to whole datasets. We repeat our suggestion algorithm for 128 times and record the number of correct suggestions according to our empirical knowledge. The results (in Table \ref{tab:mc_correctness}) show that our algorithm suggests the correct $\tau$ values in few seconds, and achieves higher than 90\% accuracy with only 100 samples for each iteration.

\section{Related work}
\label{sec:relatedwork}

Most of recent works on set-similarity joins (e.g., \cite{conf/sigmod/LuLWLW13,journals/tkde/ShangLLF16}) follow the filter-and-verification framework. Verification is expensive; hence, the key technical challenge is to design a filtering mechanism to prune away irrelevant record pairs as much as possible. Several techniques have been proposed, such as length \cite{conf/icde/LiLL08}, position \cite{conf/www/XiaoWLY08}, and prefix filtering \cite{conf/www/BayardoMS07}. While the last one is widely used in many works of literature, it suffers the problem of numerous candidates since it finds all record pairs having at least one overlapped token. This problem is tackled by Wang \textit{et al.} \cite{conf/sigmod/WangLF12} with a time-complexity-based method. For our research problem, the most recent article \cite{journals/tkde/ShangLLF16} extends the prefix filtering by considering the weight of each token, but still having the same problem due to the one-overlap policy (see $\tau=1$ in Figure \ref{fig:exp_tau}). In contrast, our work finds the best overlap constraint for each dataset to achieve a much shorter joining time.

\section{Conclusion and Future work}
\label{sec:conclusion}

This paper studies a problem of integrating taxonomies for efficient set-similarity joins. We first extend the prefix filtering technique to solve the join problem efficiently, then, as our technical contribution, we propose a novel estimation framework to judiciously select the parameter which minimises the total running time. Experiments based on real datasets exhibit the superiority of proposed algorithms. As future work, we would like to apply our estimation method for accelerating the computation of knowledge-based similarity matrices used in various machine learning tasks (e.g., \cite{conf/sigir/ZhangL0ZLM14}).

\smallskip

\noindent \textbf{Acknowledgement}. This work is supported by the Academy of Finland (310321). Contact author and email: jiaheng.lu@helsinki.fi.


\bibliographystyle{ACM-Reference-Format}
\bibliography{refs}


\begin{thebibliography}{00}


\ifx \showCODEN    \undefined \def \showCODEN     #1{\unskip}     \fi
\ifx \showDOI      \undefined \def \showDOI       #1{#1}\fi
\ifx \showISBNx    \undefined \def \showISBNx     #1{\unskip}     \fi
\ifx \showISBNxiii \undefined \def \showISBNxiii  #1{\unskip}     \fi
\ifx \showISSN     \undefined \def \showISSN      #1{\unskip}     \fi
\ifx \showLCCN     \undefined \def \showLCCN      #1{\unskip}     \fi
\ifx \shownote     \undefined \def \shownote      #1{#1}          \fi
\ifx \showarticletitle \undefined \def \showarticletitle #1{#1}   \fi
\ifx \showURL      \undefined \def \showURL       {\relax}        \fi
\providecommand\bibfield[2]{#2}
\providecommand\bibinfo[2]{#2}
\providecommand\natexlab[1]{#1}
\providecommand\showeprint[2][]{arXiv:#2}

\bibitem[\protect\citeauthoryear{Arasu, Chaudhuri, and Kaushik}{Arasu
  et~al\mbox{.}}{2008}]%
        {conf/icde/ArasuCK08}
\bibfield{author}{\bibinfo{person}{Arvind Arasu}, \bibinfo{person}{Surajit
  Chaudhuri}, {and} \bibinfo{person}{Raghav Kaushik}.}
  \bibinfo{year}{2008}\natexlab{}.
\newblock \showarticletitle{{Transformation-based Framework for Record
  Matching}}. In \bibinfo{booktitle}{{\em ICDE}}. \bibinfo{pages}{40--49}.
\newblock


\bibitem[\protect\citeauthoryear{Bayardo, Ma, and Srikant}{Bayardo
  et~al\mbox{.}}{2007}]%
        {conf/www/BayardoMS07}
\bibfield{author}{\bibinfo{person}{Roberto~J Bayardo}, \bibinfo{person}{Yiming
  Ma}, {and} \bibinfo{person}{Ramakrishnan Srikant}.}
  \bibinfo{year}{2007}\natexlab{}.
\newblock \showarticletitle{{Scaling up all pairs similarity search}}. In
  \bibinfo{booktitle}{{\em WWW}}. \bibinfo{publisher}{ACM},
  \bibinfo{pages}{131--140}.
\newblock


\bibitem[\protect\citeauthoryear{Finch}{Finch}{2009}]%
        {journals/cam/finch2009}
\bibfield{author}{\bibinfo{person}{Tony Finch}.}
  \bibinfo{year}{2009}\natexlab{}.
\newblock \showarticletitle{{Incremental calculation of weighted mean and
  variance}}.
\newblock \bibinfo{journal}{{\em University of Cambridge\/}}
  \bibinfo{volume}{4} (\bibinfo{year}{2009}), \bibinfo{pages}{11--15}.
\newblock


\bibitem[\protect\citeauthoryear{Giovannoni}{Giovannoni}{2017}]%
        {giovannoni2017personalized}
\bibfield{author}{\bibinfo{person}{Gavin Giovannoni}.}
  \bibinfo{year}{2017}\natexlab{}.
\newblock \showarticletitle{{Personalized medicine in multiple sclerosis}}.
\newblock \bibinfo{journal}{{\em NDM\/}} \bibinfo{volume}{7},
  \bibinfo{number}{6s} (\bibinfo{year}{2017}), \bibinfo{pages}{13--17}.
\newblock


\bibitem[\protect\citeauthoryear{Li, Lu, and Lu}{Li et~al\mbox{.}}{2008}]%
        {conf/icde/LiLL08}
\bibfield{author}{\bibinfo{person}{Chen Li}, \bibinfo{person}{Jiaheng Lu},
  {and} \bibinfo{person}{Yiming Lu}.} \bibinfo{year}{2008}\natexlab{}.
\newblock \showarticletitle{Efficient Merging and Filtering Algorithms for
  Approximate String Searches}. In \bibinfo{booktitle}{{\em {ICDE}}}.
  \bibinfo{pages}{257--266}.
\newblock


\bibitem[\protect\citeauthoryear{Lu, Lin, Wang, Li, and Wang}{Lu
  et~al\mbox{.}}{2013}]%
        {conf/sigmod/LuLWLW13}
\bibfield{author}{\bibinfo{person}{Jiaheng Lu}, \bibinfo{person}{Chunbin Lin},
  \bibinfo{person}{Wei Wang}, \bibinfo{person}{Chen Li}, {and}
  \bibinfo{person}{Haiyong Wang}.} \bibinfo{year}{2013}\natexlab{}.
\newblock \showarticletitle{{String similarity measures and joins with
  synonyms}}. In \bibinfo{booktitle}{{\em SIGMOD}}. \bibinfo{publisher}{ACM},
  \bibinfo{pages}{373--384}.
\newblock


\bibitem[\protect\citeauthoryear{Lu, Lin, Wang, Li, and Xiao}{Lu
  et~al\mbox{.}}{2015}]%
        {journals/tods/LuLWLX15}
\bibfield{author}{\bibinfo{person}{Jiaheng Lu}, \bibinfo{person}{Chunbin Lin},
  \bibinfo{person}{Wei Wang}, \bibinfo{person}{Chen Li}, {and}
  \bibinfo{person}{Xiaokui Xiao}.} \bibinfo{year}{2015}\natexlab{}.
\newblock \showarticletitle{Boosting the Quality of Approximate String Matching
  by Synonyms}.
\newblock \bibinfo{journal}{{\em {ACM} Trans. Database Syst.\/}}
  \bibinfo{volume}{40}, \bibinfo{number}{3} (\bibinfo{year}{2015}),
  \bibinfo{pages}{15:1--15:42}.
\newblock


\bibitem[\protect\citeauthoryear{Munkres}{Munkres}{1957}]%
        {journals/JSIAM/Munkres57}
\bibfield{author}{\bibinfo{person}{J Munkres}.}
  \bibinfo{year}{1957}\natexlab{}.
\newblock \showarticletitle{{Algorithms for the Assignment and Transportation
  Problems}}.
\newblock \bibinfo{journal}{{\em JSIAM\/}} \bibinfo{volume}{5},
  \bibinfo{number}{1} (\bibinfo{year}{1957}), \bibinfo{pages}{32--38}.
\newblock


\bibitem[\protect\citeauthoryear{Sarawagi and Kirpal}{Sarawagi and
  Kirpal}{2004}]%
        {conf/sigmod/Sarawagi04}
\bibfield{author}{\bibinfo{person}{Sunita Sarawagi} {and} \bibinfo{person}{Alok
  Kirpal}.} \bibinfo{year}{2004}\natexlab{}.
\newblock \showarticletitle{{Efficient set joins on similarity predicates}}. In
  \bibinfo{booktitle}{{\em SIGMOD}}. \bibinfo{pages}{743--754}.
\newblock


\bibitem[\protect\citeauthoryear{Shang, Liu, Li, and Feng}{Shang
  et~al\mbox{.}}{2016}]%
        {journals/tkde/ShangLLF16}
\bibfield{author}{\bibinfo{person}{Zeyuan Shang}, \bibinfo{person}{Yaxiao Liu},
  \bibinfo{person}{Guoliang Li}, {and} \bibinfo{person}{Jianhua Feng}.}
  \bibinfo{year}{2016}\natexlab{}.
\newblock \showarticletitle{{K-Join: Knowledge-Aware Similarity Join}}.
\newblock \bibinfo{journal}{{\em TKDE\/}} \bibinfo{volume}{28},
  \bibinfo{number}{12} (\bibinfo{year}{2016}), \bibinfo{pages}{3293--3308}.
\newblock


\bibitem[\protect\citeauthoryear{Wang, Li, and Feng}{Wang
  et~al\mbox{.}}{2012}]%
        {conf/sigmod/WangLF12}
\bibfield{author}{\bibinfo{person}{Jiannan Wang}, \bibinfo{person}{Guoliang
  Li}, {and} \bibinfo{person}{Jianhua Feng}.} \bibinfo{year}{2012}\natexlab{}.
\newblock \showarticletitle{{Can we beat the prefix filtering?: an adaptive
  framework for similarity join and search}}. In \bibinfo{booktitle}{{\em
  SIGMOD}}. \bibinfo{pages}{85--96}.
\newblock


\bibitem[\protect\citeauthoryear{Xiao, Wang, Lin, and Yu}{Xiao
  et~al\mbox{.}}{2008}]%
        {conf/www/XiaoWLY08}
\bibfield{author}{\bibinfo{person}{Chuan Xiao}, \bibinfo{person}{Wei Wang},
  \bibinfo{person}{Xuemin Lin}, {and} \bibinfo{person}{Jeffrey~Xu Yu}.}
  \bibinfo{year}{2008}\natexlab{}.
\newblock \showarticletitle{{Efficient similarity joins for near duplicate
  detection}}. In \bibinfo{booktitle}{{\em WWW}}. \bibinfo{publisher}{ACM},
  \bibinfo{pages}{131--140}.
\newblock


\bibitem[\protect\citeauthoryear{Xu and Lu}{Xu and Lu}{2017}]%
        {conf/dasfaa/XuL17}
\bibfield{author}{\bibinfo{person}{Pengfei Xu} {and} \bibinfo{person}{Jiaheng
  Lu}.} \bibinfo{year}{2017}\natexlab{}.
\newblock \showarticletitle{Top-k String Auto-Completion with Synonyms}. In
  \bibinfo{booktitle}{{\em {DASFAA}}}. \bibinfo{pages}{202--218}.
\newblock


\bibitem[\protect\citeauthoryear{Zhang, Lai, Zhang, Zhang, Liu, and Ma}{Zhang
  et~al\mbox{.}}{2014}]%
        {conf/sigir/ZhangL0ZLM14}
\bibfield{author}{\bibinfo{person}{Yongfeng Zhang}, \bibinfo{person}{Guokun
  Lai}, \bibinfo{person}{Min Zhang}, \bibinfo{person}{Yi Zhang},
  \bibinfo{person}{Yiqun Liu}, {and} \bibinfo{person}{Shaoping Ma}.}
  \bibinfo{year}{2014}\natexlab{}.
\newblock \showarticletitle{{Explicit factor models for explainable
  recommendation based on phrase-level sentiment analysis}}. In
  \bibinfo{booktitle}{{\em SIGIR}}. \bibinfo{publisher}{ACM},
  \bibinfo{pages}{83--92}.
\newblock


\end{thebibliography}

\end{document}